\documentclass{PoS}
\usepackage{amsfonts}

\font\twelveeusm=eusm10 scaled 1200
\font\teneusm=eusm10
  \newfam\eusmfam
  \def\eusm{\fam\eusmfam\twelveeusm}
  \textfont\eusmfam=\twelveeusm
  \scriptfont\eusmfam=\teneusm
  \scriptscriptfont\eusmfam=\scriptfont\eusmfam
\font\twelvefrak=eufm10 scaled 1200
\font\tenfrak=eufm10
  \newfam\frakfam
  
  \textfont\frakfam=\twelvefrak
  \scriptfont\frakfam=\tenfrak
  \scriptscriptfont\frakfam=\scriptfont\frakfam
\def\sqr#1#2{{\vcenter{\hrule height.#2pt
   \hbox{\vrule width.#2pt height#1pt \kern#1pt
      \vrule width.#2pt}
   \hrule height.#2pt}}}

\def\bsqr#1#2{{\vrule width #1pt height#2pt}}
\def\bsquare{{\mathchoice\bsqr66\bsqr66\bsqr33\bsqr33}}
\def\badbreak{\penalty1000}

\title{Deconfinement, Chiral Symmetry Breaking and Chiral Polarization}

\ShortTitle{Deconfinement, Chiral Symmetry Breaking\ldots}

\author{Andrei Alexandru\\
        The George Washington University, Washington, DC, USA\\
        E-mail: \email{aalexan@gwu.edu}}

\author{\speaker{Ivan Horv\'ath}%
         \thanks{We are indebted to Mingyang Sun for providing some of the illustrative graphics.}\\
        University of Kentucky, Lexington, KY, USA\\
        E-mail: \email{horvath@pa.uky.edu}}

\abstract{We examine the feasibility of the proposition that there is a temperature range 
 T$_c$ < T < T$_{ch}$ in N$_f$=0 QCD, where real Polyakov line (deconfined) vacuum exhibits 
 valence spontaneous chiral symetry breaking and dynamical chiral polarization of 
 Dirac eigenmodes. Detailed finite--volume analysis convincingly demonstrates the existence 
 of such phase at fixed cutoff (a=0.085 fm). Moreover, it is found that this behavior also
 takes place closer to the continuum limit (a=0.060 fm) without qualitative change in its 
 properties.}  

\FullConference{The 32nd International Symposium on Lattice Field Theory,\\
                23-28 June, 2014\\
                Columbia University New York, NY}

\begin{document}

\noindent {\bf 1. The Context.} This presentation is concerned with two proposed features
in SU(3) gauge theories that closely relate to Dirac eigenmodes. The first one 
suggests~\cite{Edw99A} that, when N$_f$=0 system deconfines into ``real Polyakov line'' 
vacuum at $T_c$, {\em valence} chiral symmetry remains broken for some range of temperatures 
T$_c$ < T < T$_{ch}$. Only after crossing such {\em anomalous} phase does the system 
become both deconfined and chirally symmetric. 

\begin{figure}[b]
\begin{center}
    \vskip -0.05in
    \centerline{
    \hskip -0.00in
    \includegraphics[height=4.5truecm,angle=0]{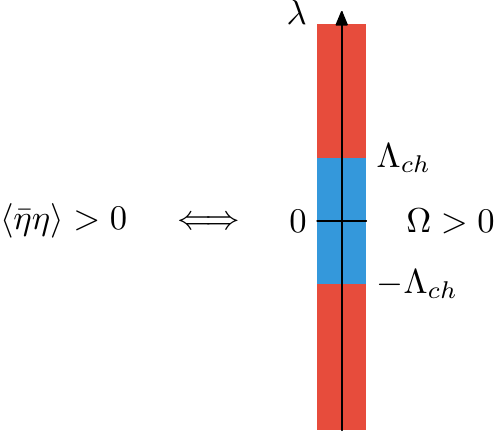}
    \hskip 1.50in
    \includegraphics[width=4.5truecm,angle=0]{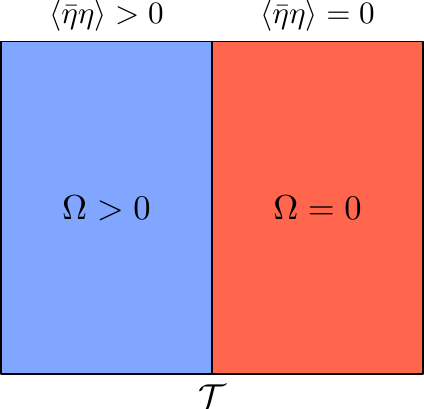}
     }
     \vskip -0.00in
     \caption{Schematic representation of vSChSB--ChP correspondence: $\eta$ denotes massless 
     valence field, $\Omega$ volume density of chirally polarized modes, and $\Lambda_{ch}$ 
     chiral polarization scale. Set ${\eusm T}$ consists of SU(3) gauge theories with N$_f$ 
     arbitrarily massive fundamental fermions at arbitrary temperature.}
     \label{fig:vschsb_chp}
     \vskip -0.0in 
\end{center}
\end{figure}

The second feature proposes that SU(3) gauge theory that breaks valence chiral symmetry 
supports a layer of chirally polarized Dirac modes at low end of the spectrum~\cite{Ale12D,Ale14A}, 
as shown schematically in Fig.~\ref{fig:vschsb_chp} (left).
Such layer is only well--defined via {\em dynamical} polarization methods~\cite{Ale10A}, 
and its existence is indicated e.g. by volume density $\Omega$ of participating modes. 
The above relationship is conjectured to be an equivalence: splitting SU(3) theories 
into chirally broken and chirally symmetric is to yield the same result as partition into 
chirally polarized and chirally anti--polarized cases 
(see Fig.~\ref{fig:vschsb_chp}(right) for details). Note that $\eta$ is a massless valence 
field and the question of symmetry breakdown is well--defined with dynamical quarks of arbitrary 
masses.

The above proposals are each interesting in their own right. Indeed, the real Polyakov line 
vacuum of deconfined N$_f$=0 theory is a good qualitative proxy to gauge vacuum of 
``real world'' QCD above finite--temperature transition. It is thus important to demonstrate 
that the anomalous phase is a true feature of N$_f$=0 theory, i.e. that of infinite--volume 
continuum system. 
In the other case, the validity of chiral polarization connection significantly 
constrains possible mechanisms of spontaneous chiral symmetry breaking for SU(3) 
theories in a model--independent way. After all, chiral condensate and chiral polarization 
are not required to be tied together by general principles. 

However, the main reason we brought the above two aspects together here is that the former 
provides for a particularly sensitive test of the latter: is valence spontaneous chiral 
symmetry breaking (vSChSB)$\,$--$\,$chiral polarization (ChP) correspondence satisfied in 
corners of SU(3) theory space that are in anomalous phase?

\smallskip
\noindent {\bf 2. The Tools.} To detect vSChSB we use the fact that this phenomenon is 
equivalent to Dirac mode condensation. As such, it is indicated by non--zero density
of near--zero modes $\rho(\lambda\to 0)$ in the infinite volume limit. The presence
of chiral polarization will be signaled by volume density $\Omega$ of polarized modes, 
on whose definition we now briefly elaborate.

The concept of dynamical ChP is based on a correlation coefficient $C_A$ of Ref.~\cite{Ale10A},
uniquely associated with property of polarization. Positive correlation implies tendency 
for asymmetry in left and right (chiral polarization), while the negative correlation signifies 
tendency for symmetry (chiral anti--polarization). Since $C_A$ can be assigned to every mode, 
it is useful to define the associated spectral constructs~\cite{Ale12D,Ale14A}. 
For example, cumulative polarization density $\sigma_{ch}(\lambda)$ is considered in 
addition to the usual cumulative mode density $\sigma(\lambda)$. In particular
\begin{equation}
  \sigma_{ch}(\lambda,V) \equiv \frac{1}{V} \, 
    \langle \, \sum_{0 < \lambda_k < \lambda} \, C_{A,k}\;\rangle
  \qquad\qquad
  \sigma(\lambda,V) \equiv \frac{1}{V} \, 
  \langle \, \sum_{0 < \lambda_k < \lambda} \, 1\;\rangle
\end{equation}
where $\langle \ldots \rangle$ denotes the ensemble average. Exact zero modes are not
included since their contribution vanishes in thermodynamic limit. In chirally polarized 
theory, $\sigma_{ch}(\lambda)$ will grow away from zero until it hits the edge of chirally 
polarized layer $\Lambda_{ch}$. The value of $\sigma$ at this point is $\Omega$.

\begin{figure}[b]
\begin{center}
    \vskip -0.05in
    \centerline{
    \hskip -0.00in
    \includegraphics[width=15.0truecm,angle=0]{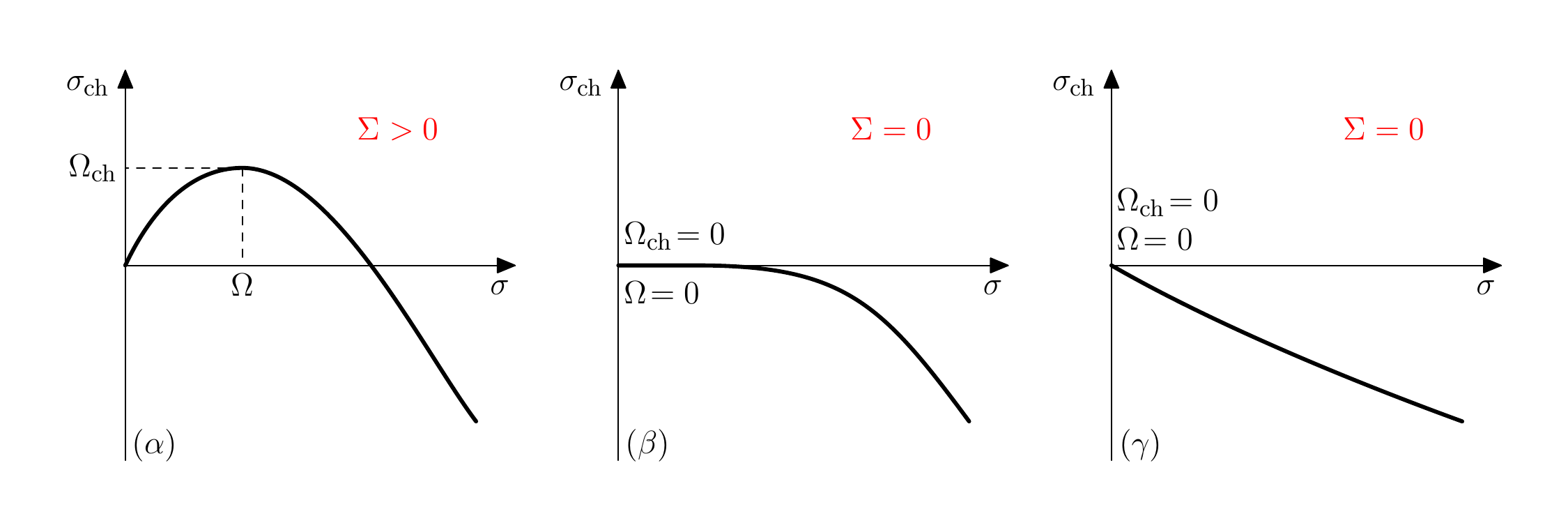}
     }
     \vskip -0.00in
     \caption{Behaviors of $\sigma_{ch}(\sigma)$ in the vicinity of $\sigma=0$. 
     vSChSB--ChP correspondence associates case $(\alpha)$ with spontaneously 
     broken valence chiral symmetry while cases $(\beta)$ and $(\gamma)$ with
     symmetric vacuum.}
     \label{fig:illus_2}
     \vskip -0.0in 
\end{center}
\end{figure}

It is convenient to eliminate $\lambda$ from the above relationships and consider 
$\sigma_{ch}=\sigma_{ch}(\sigma)$~\cite{Ale14A}. Here too, chiral polarization 
reveals itself as a positive bump in the vicinity of the origin, defining 
$\Omega$ and volume density of total chirality $\Omega_{ch}$ simultaneously 
(see Fig.~\ref{fig:illus_2}). If $\sigma_{ch}(\sigma)$ turns negative away from 
$\sigma=0$, or remains zero, there is no chirally polarized layer and vSChSB--ChP 
correspondence predicts that valence chiral symmetry will not be broken in the
infinite--volume limit.

All results presented here relate to Wilson gauge action and scale set via $r_0=0.5$ fm. 
We use overlap Dirac operator~\cite{Neu98BA} ($\rho=26/19$) with Wilson--Dirac kernel 
($r=1$) as a massless valence probe. As emphasized already, our interest is in deconfined 
vacuum with real Polyakov line.

\smallskip
\noindent{\bf 3. Qualitative Overview.} To start this inquiry, we seek to reproduce 
the qualitative observation in Ref.~\cite{Edw99A} of anomalous accumulation of 
overlap near--zeromodes past deconfinement temperature $T_c$, and to check whether 
their chiral polarization properties concur with vSChSB--ChP correspondence. Using 
the fixed scale approach on $20^3\times$ N$_t$ lattice,
we change temperature across $T_c$. With scale set at $\,a=0.085$ fm ($\beta=6.054$), 
the fixed 3--volume is (1.7 fm)$^3$ and the nominal continuum--extrapolated 
deconfinement transition (see Ref.~\cite{Kar97A}) occurs just above N$_t$=8.

\begin{figure}[t]
\begin{center}
    \vskip -0.10in
    \centerline{
    \hskip 0.10in
    \includegraphics[height=3.7truecm,angle=0]{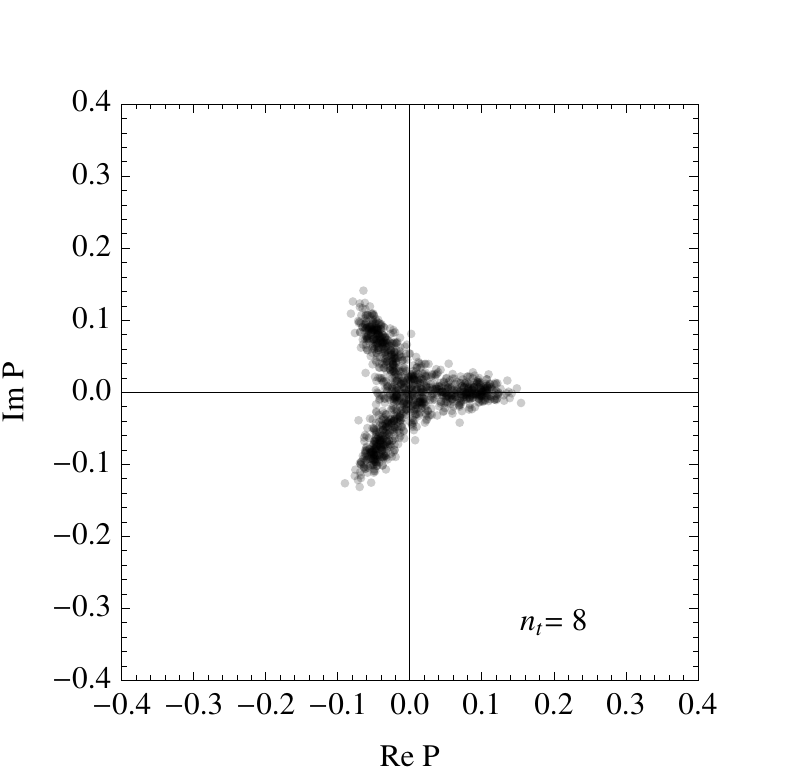}
    \hskip -0.00in
    \includegraphics[width=5.5truecm,angle=0]{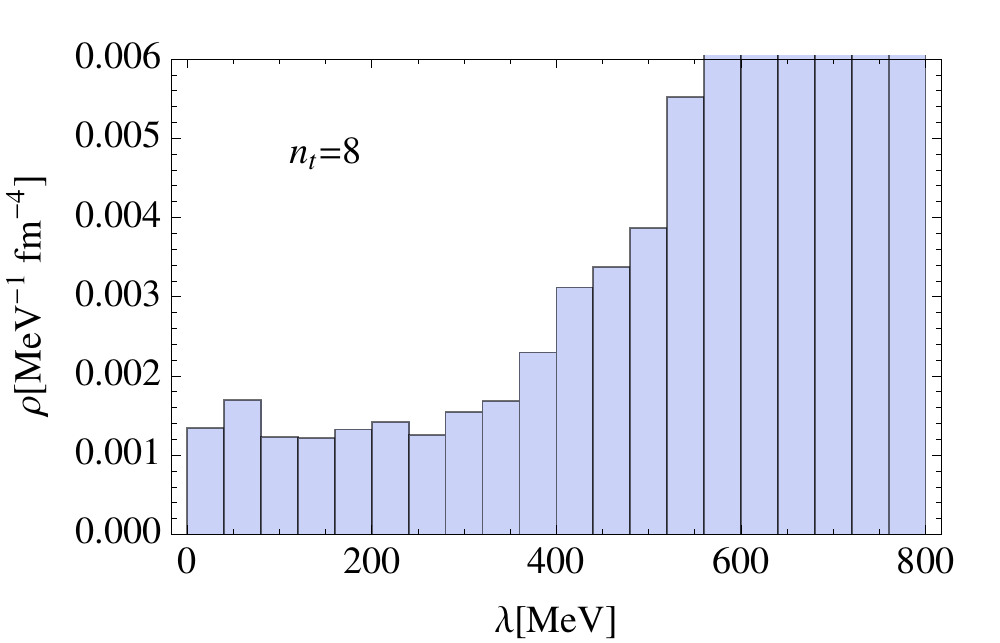}
    \hskip -0.00in
    \includegraphics[width=5.5truecm,angle=0]{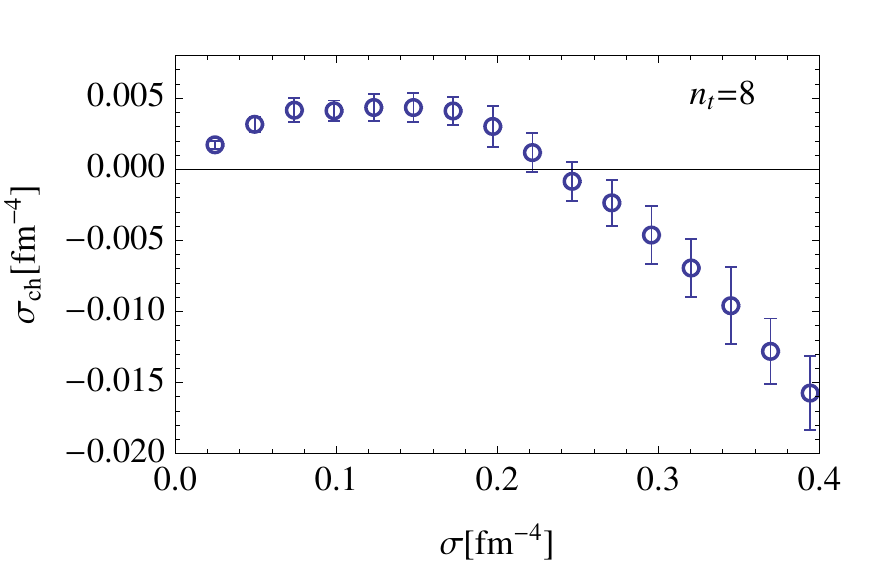}
     }
     \vskip -0.00in
    \centerline{
    \hskip 0.10in
    \includegraphics[height=3.7truecm,angle=0]{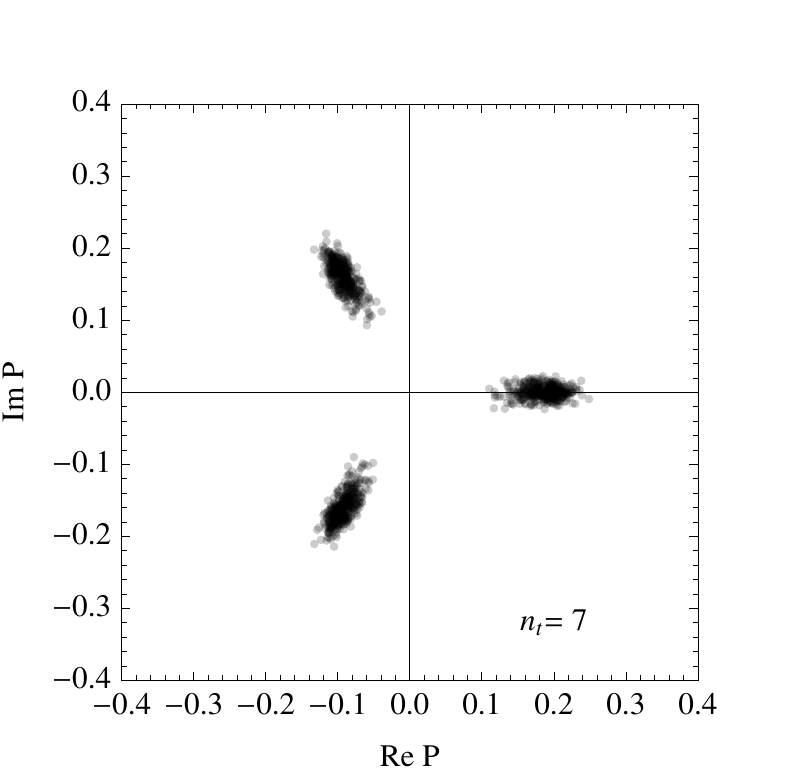}
    \hskip -0.00in
    \includegraphics[width=5.5truecm,angle=0]{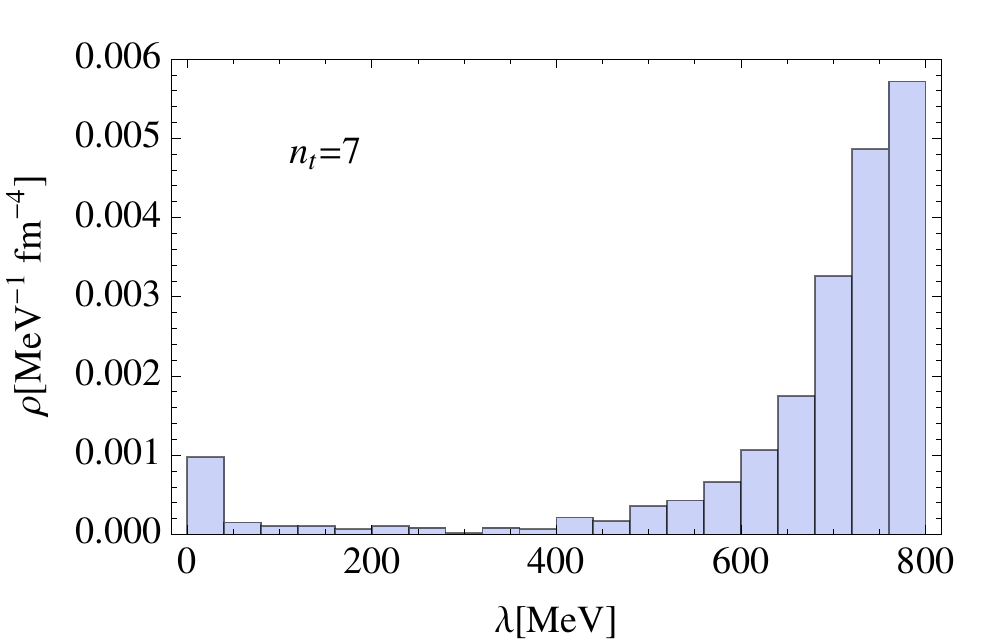}
    \hskip -0.00in
    \includegraphics[width=5.5truecm,angle=0]{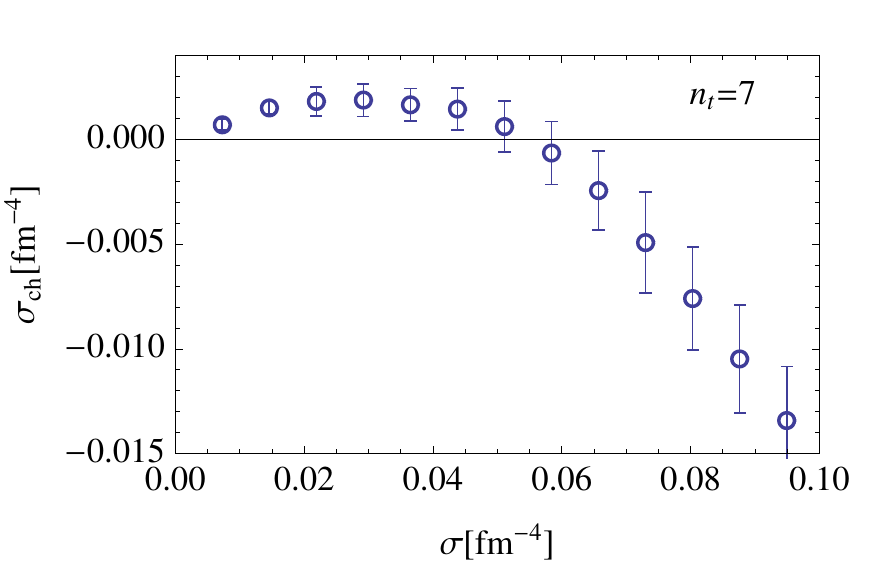}
     }
     \vskip -0.00in
    \centerline{
    \hskip 0.10in
    \includegraphics[height=3.7truecm,angle=0]{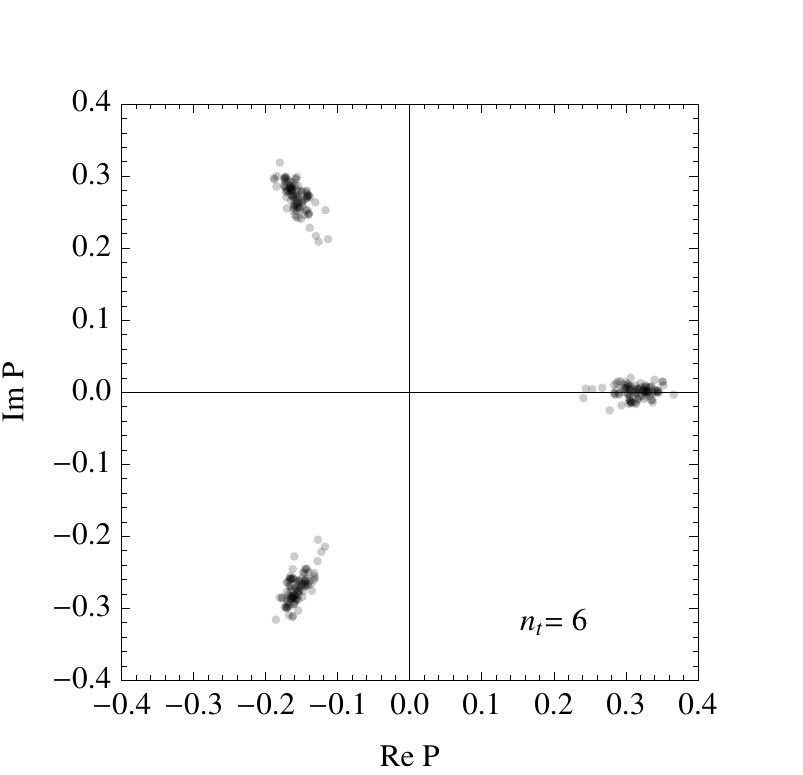}
    \hskip -0.00in
    \includegraphics[width=5.5truecm,angle=0]{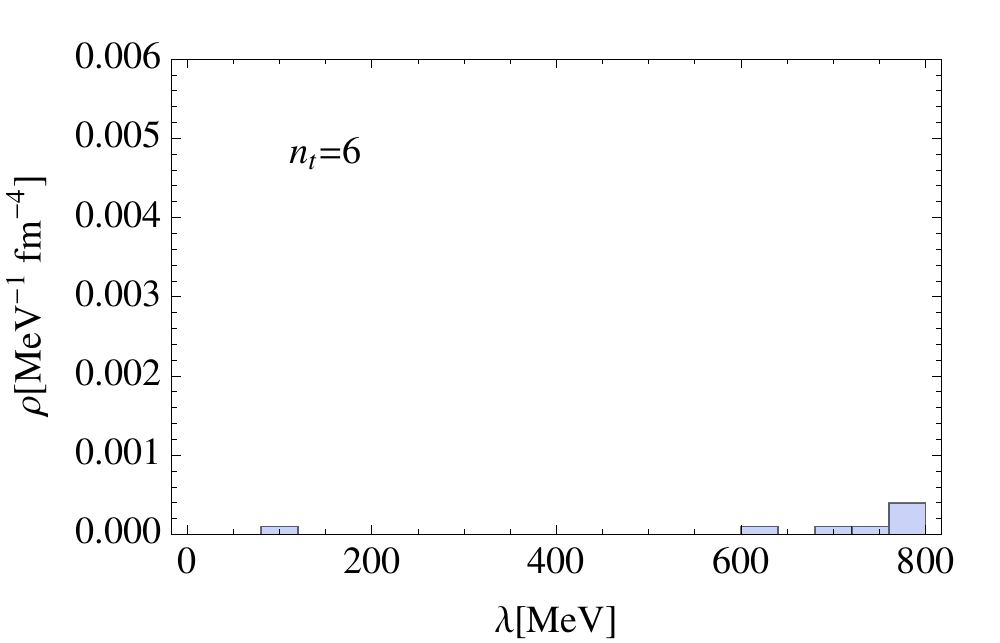}
    \hskip -0.00in
    \includegraphics[width=5.5truecm,angle=0]{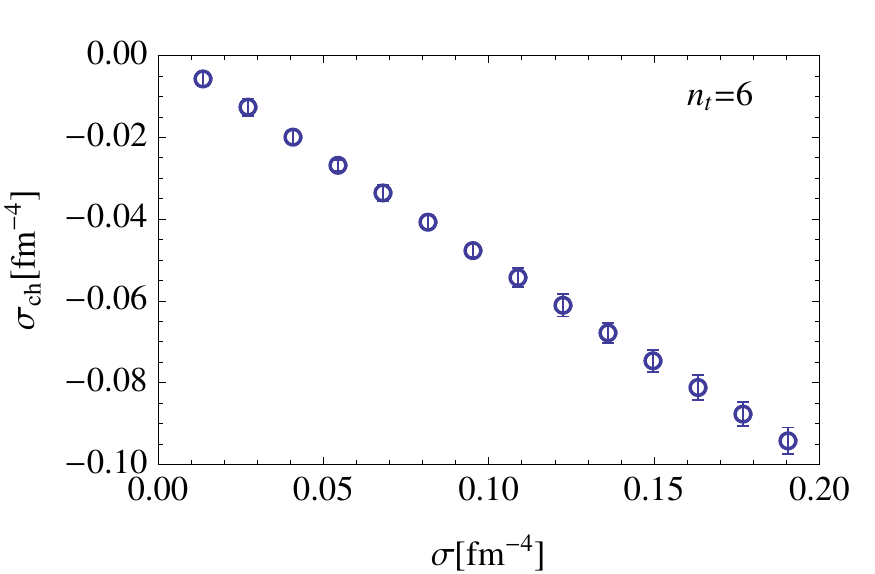}
     }
     \vskip -0.00in
     \caption{Indicators of deconfinement (left column), valence chiral symmetry breaking 
     (middle column) and chiral polarization (right column) on 20$^3\times$ N$_t$ lattices. 
     See discussion in the text.}
     \label{fig:qualit}
     \vskip -0.35in 
\end{center}
\end{figure}

While detailed account of these calculations covering wide range of temperatures is 
given elsewhere~\cite{Ale12D,Ale14A}, Fig.~\ref{fig:qualit} shows the relevant overview 
results for N$_t$=8,7,6. Scatter plots of the Polyakov line in left column confirm 
that N$_t$=8 system behaves as borderline--deconfined, while at N$_t$=7 and 6 it is 
manifestly deconfined. Nevertheless, the mode density data in the second column shows
abundant near--zeromodes at N$_t$=7 of anomalous type (non-monotonic $\rho(\lambda)$) 
observed in~\cite{Edw99A}. These modes disappear at N$_t$=6, suggesting valence 
chiral transition between N$_t$=7 and 6. In the third column of Fig.~\ref{fig:qualit}, 
the behavior of $\sigma_{ch}(\sigma)$ clearly indicates the onset of chiral 
anti--polarization at the same place, in accordance with vSChSB--ChP correspondence.

\smallskip
\noindent{\bf 3. Fixed Cutoff.} Since symmetry breaking is involved in both features studied
here, we are dealing with infinite--volume propositions, and finite--volume considerations 
are important. In particular, it is essential to truly establish the existence of deconfined 
chirally broken dynamics on the lattice, i.e. to carefully study volume trends at fixed 
cutoff~\cite{Ale14A}. One explicit worry is that the observed anomalous behavior could be 
an unwanted feature of the overlap construction that may be ``cured'' by large volumes 
in a way similar to the exceptional configuration issue.  

\begin{figure}[t]
\begin{center}
    \vskip -0.05in
    \centerline{
    \hskip -0.00in
    \includegraphics[width=6.5truecm,angle=0]{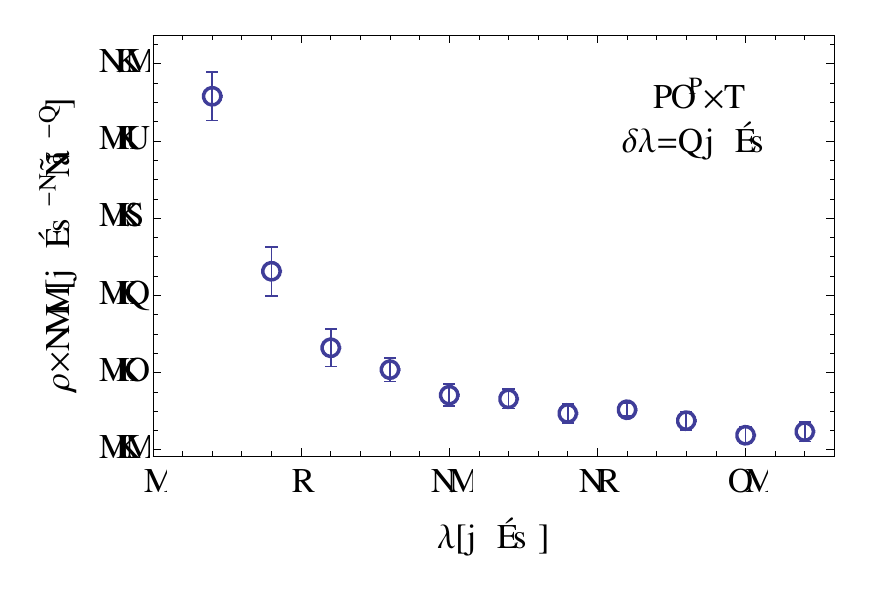}
    \hskip 0.1in
    \includegraphics[width=7.0truecm,angle=0]{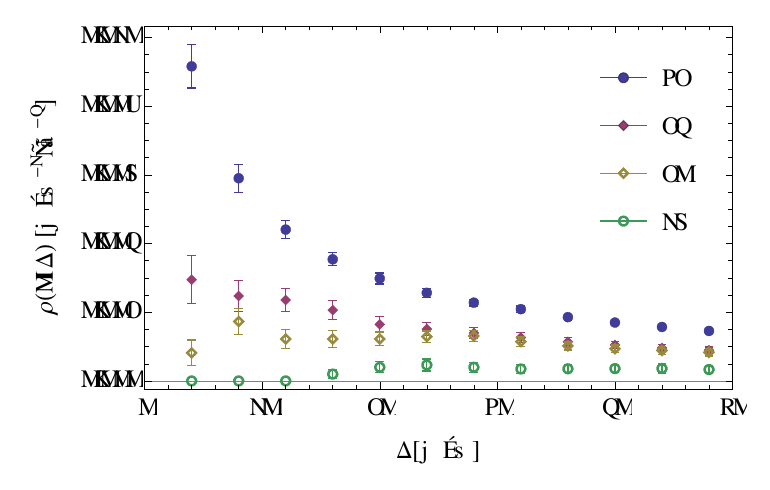}
     }
     \vskip -0.00in
     \caption{Spectral mode density in the vicinity of origin at N$_t$=7 for the largest 
     volume (left) and the mode (chiral) condensation parameter as a function of 
     coarse--graining range $\Delta$ (right).}
     \label{fig:rho_vol}
     \vskip -0.30in 
\end{center}
\end{figure}

To examine this, we study the N$_t$=7 theory which clearly offers itself as a candidate of 
dynamics in anomalous phase. Thus, N$^3 \times$7 systems are considered at the previously 
mentioned gauge coupling, and N=16,20,24,32. The largest volume in the sequence is thus 
(2.72 fm)$^3$. The closeup of its spectral density at low end of the spectrum is provided 
in Fig.~\ref{fig:rho_vol} (left), clearly confirming the anomalous accumulation 
of near--zero modes. 
To check volume tendencies, we consider a coarse--grained mode condensation parameter 
$\rho(\lambda=0,\Delta,V) \equiv \sigma(\lambda=\Delta,V)/\Delta$. Infinite--volume limit at 
fixed coarse--graining parameter $\Delta$ needs to be taken before $\Delta \to 0$ defines 
the condensate. Fig.~\ref{fig:rho_vol} (right) shows $\rho(0,\Delta)$ for above volumes 
down to $\Delta = 4$ MeV. Essential feature of these results is that 
$\rho(0,\Delta,V)$ grows as a function of volume at arbitrary accessible $\Delta$: larger 
volumes enhance the density of near--zero modes. Moreover, the rate of increase grows 
with decreasing $\Delta$ making it extremely unlikely that the mode (chiral) condensate 
vanishes, and confirming that $\rho(\lambda)$ exhibits anomalous (monotonically decreasing) 
behavior near the origin in the infinite volume limit. 

\begin{figure}[b]
\begin{center}
    \vskip -0.05in
    \centerline{
    \hskip -0.00in
    \includegraphics[width=13.0truecm,angle=0]{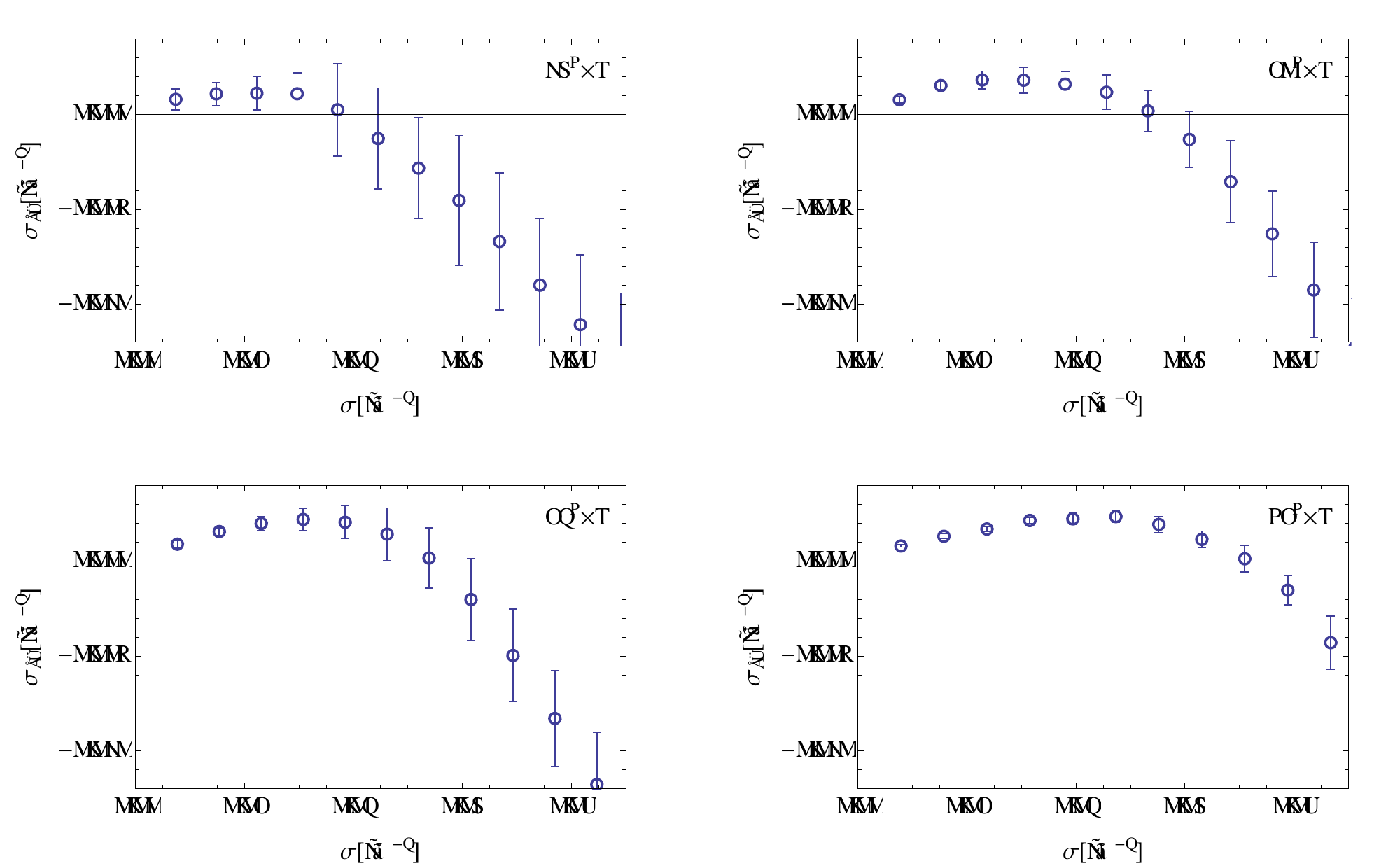}
     }
     \vskip -0.00in
     \caption{Behavior of $\sigma_{ch}(\sigma)$ at N$_t$=7: degree of chiral polarization 
     increases with increasing volume.}
     \label{fig:sigmach_vol}
     \vskip -0.30in 
\end{center}
\end{figure}

Given the above result, if vSChSB--ChP correspondence is obeyed by the regularized system 
in question, chiral polarization has to be present at least in arbitrarily large finite 
volume (see {\em Conjecture 3} of Ref.~\cite{Ale14A}). To check for this, we plot 
$\sigma_{ch}(\sigma)$ for different volumes in Fig.~\ref{fig:sigmach_vol}. As one can see,
not only is the system chirally polarized in each casee, the degree of chiral 
polarization increases as the system grows. For example, the density of chirally polarized
modes $\Omega$ (position of the maximum in $\sigma_{ch}(\sigma)$) monotonically 
increases with volume~\cite{Ale14A}. 

\smallskip
\noindent{\bf 4. Continuum Limit.} Another very relevant concern regarding the reality
of chirally broken deconfined phase is that the observed anomalous behavior might be 
due to lattice artifacts that will disappear sufficiently close to the continuum limit. To 
examine this possibility, we consider the same system at significantly larger cutoff, where 
``the same'' refers to identical $T/T_c$. Since $r_0$ serves to determine the scale, it is 
convenient to use the universal continuum--extrapolated result $r_0 T_c = 0.7498(50)$ of 
Ref.~\cite{Nec03A} to fix $T/T_c$. The systems of the previous section were at 
$a/r_0=0.1701$, which for $N_t=7$ translates into $T/T_c=1.12$. We now switch to 
$a/r_0=0.1190$ and $N_t=10$ leading to the same $T/T_c$ but with lattice spacing 
lowered to $a=0.0595$ fm. 

\begin{figure}[b]
\begin{center}
    \vskip -0.10in
    \centerline{
    \hskip -0.00in
    \includegraphics[width=7.5truecm,angle=0]{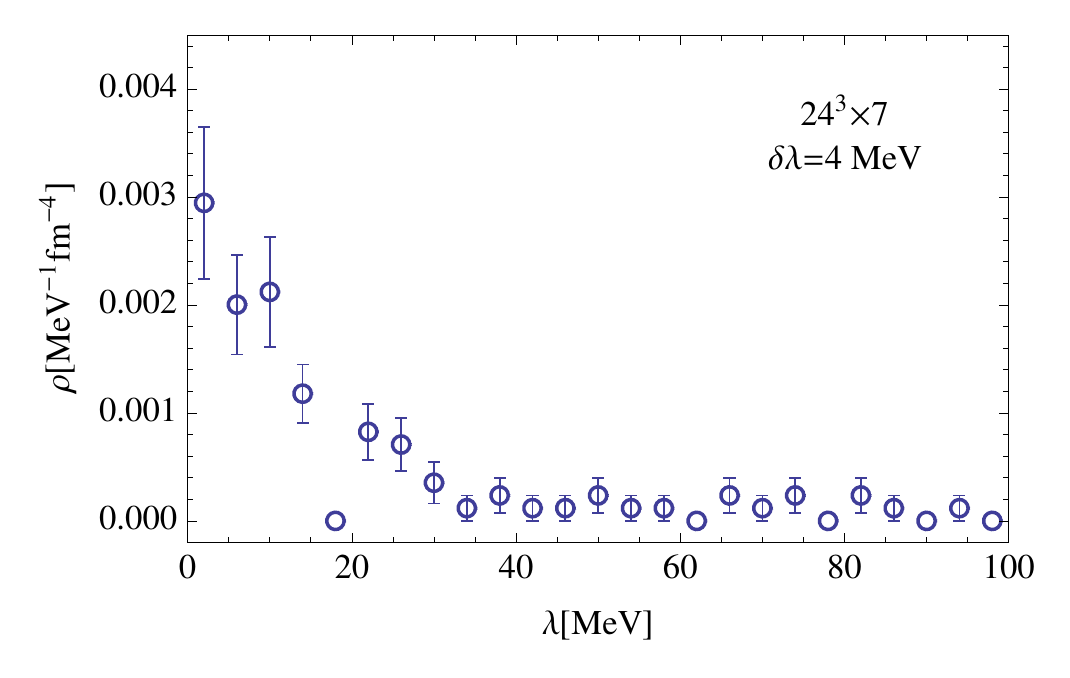}
    \hskip -0.00in
    \includegraphics[width=7.5truecm,angle=0]{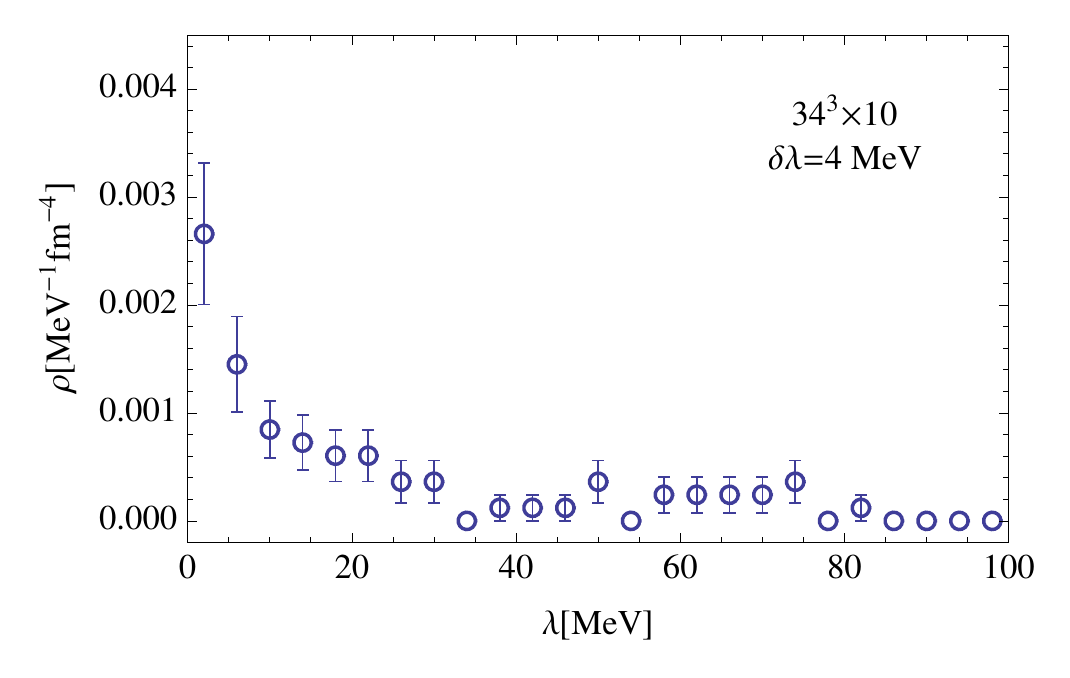}
     }
     \vskip -0.00in
    \centerline{
    \hskip -0.00in
    \includegraphics[width=7.5truecm,angle=0]{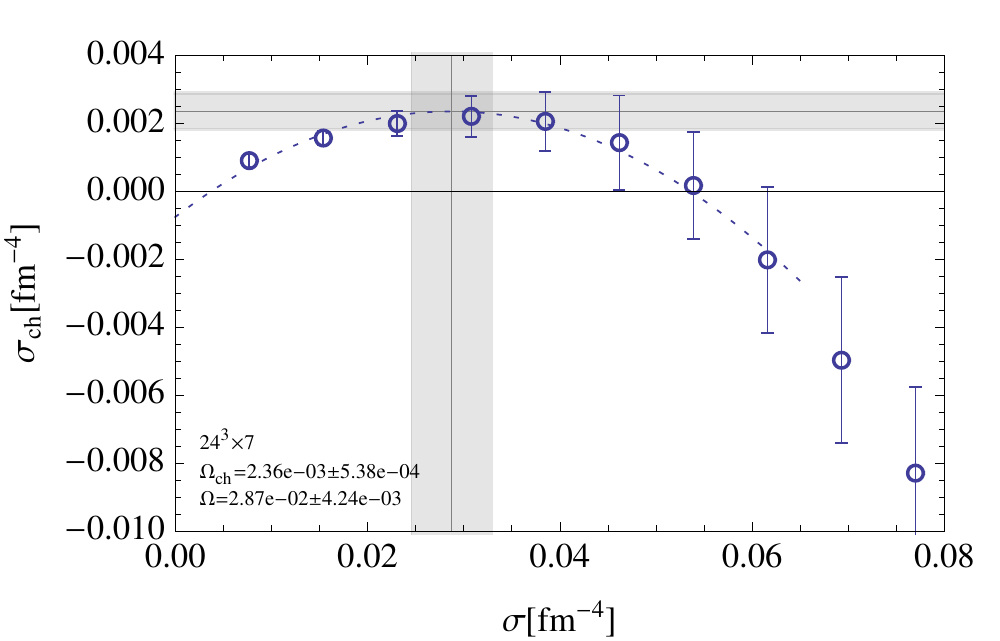}
    \hskip  0.05in
    \includegraphics[width=7.5truecm,angle=0]{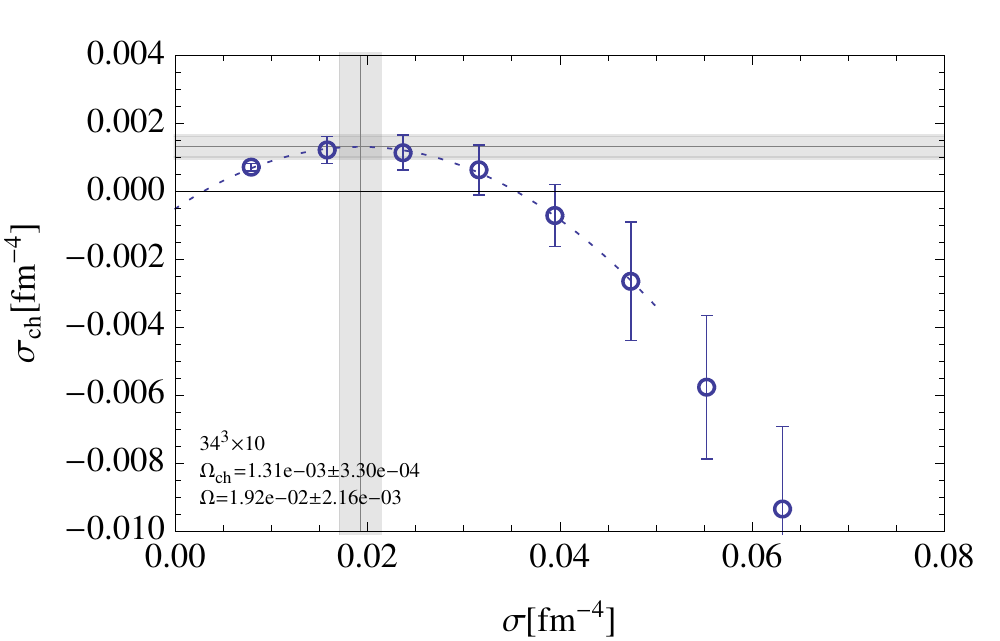}
     }
     \vskip -0.00in
     \caption{Comparison of spectral density close to zero (top) and cumulative chiral 
     polarization $\sigma_{ch}(\sigma)$ (bottom) for systems at $T/T_c=1.12$, with the same
     volume (L=2 fm), but with $a=0.0850$ fm (left) and $a=0.0595$ fm (right).}
     \label{fig:continuum}
     \vskip -0.35in 
\end{center}
\end{figure}

Using the above setup we study the $34^3 \times 10$ lattice which entails the same spatial
volume as the $24^3 \times 7$ system of the previous section (L=2.0 fm) to facilitate
the most direct comparison. In Fig.~\ref{fig:continuum} (top) we show spectral density 
$\rho(\lambda)$ at low end of the spectrum for the two cases. As one can see, the anomalous 
behavior is clearly present at higher cutoff as well. While there is some reduction in 
the width of the peak, there is no observable change in its height, i.e. the density in 
the vicinity of the origin scales within the available statistics (400 configurations). 

Moreover, the plots of $\sigma_{ch}(\sigma)$ at the bottom of Fig.~\ref{fig:continuum} 
reveal that the theory also remains chirally polarized. Combined with the type of volume 
scaling observed in the previous section, this leads us to conclude that the existence of 
anomalous phase and vSChSB--ChP correspondence both hold on the fine lattice. Also, while 
some decrease in $\Omega$ can be observed at larger cutoff, this is mainly related to 
the aforementioned reduction in the width of the anomalous peak rather than to changing 
polarization properties of the relevant modes as continuum limit is approached.

\smallskip
\noindent{\bf 5. Discussion.} Resolving the existence of anomalous phase $T_c < T < T_{ch}$
in N$_f$=0 QCD is important for understanding the nature of QCD transition region. While 
only an approximation to the ``real world'', the long history of quenched simulations 
shows that the system retains most qualitative features of full QCD, especially  
those related to gauge fluctuations. This applies to vSChSB which probes correlations of 
the {\em gauge vacuum} leading to symmetry breakdown in external fields interacting with 
it~\cite{Ale14A}.  Thus, if ascertained to exist for N$_f$=0 in real Polyakov line vacuum, 
the occurrence of anomalous phase in QCD with dynamical quarks would not be surprising. 
Some indications of such behavior have in fact been reported 
(see e.g. Refs.~\cite{Buc13A,Sha13A,Kot14A}). 

However, the whole scenario could well be just a distortion due to infrared cutoff
being too high or ultraviolet cutoff being too small. One warning is that the anomalous 
behavior is only apparent with overlap--type fermions. While one naturally tends 
to assume that chiral formulation is more reliable, it is also possible that overlap is 
``too sensitive'' to chirality and topology in this case, and simply displays artifacts 
that other formulations cannot see. Thus, it is the {\em reality} of deconfined and chirally 
broken phase that is our main focus here and, due to the non--trivial nature of involved 
extrapolations, this can be most realistically settled in N$_f$=0 theory.

We presented some of the data, first published in Ref.~\cite{Ale14A}, that convincingly 
show stability of the phenomenon in the infrared: there indeed exists {\em lattice}
N$_f$=0 theory that is simultaneously deconfined and chirally broken. In addition, we 
analyzed previously unpublished data indicating that a substantial increase in lattice 
cutoff doesn't change the situation appreciably. Thus, while additional cutoff levels 
will be examined to make final conclusions, data available at present is consistent with 
the phase persisting to the continuum limit.

In the same vein, the data presented here supports the validity of vSChSB--ChP 
correspondence for physics in the anomalous phase. Since the nature of underlying
dynamics changes substantially across deconfinement transition, this is one of the crucial
tests for the emerging association between chiral symmetry breaking and chiral polarization.

\end{document}